# Characterization of poly- and single-crystal uranium-molybdenum alloy thin films

A M Adamska[*], R Springell, T B Scott

*Interface Analysis Centre, HH Wills Physics Laboratory, University of Bristol, Tyndall Avenue, Bristol, BS8 1TL, United Kingdom*

*am.adamska@bristol.ac.uk

**Abstract**

Poly- and single-crystal thin films of U-Mo alloys have been grown both on glass and sapphire substrates by UHV magnetron sputtering. X-ray and Electron Backscatter Diffraction data indicate that for single-crystal $U_{1-x}Mo_x$ alloys, the pure cubic uranium γ-phase exists for $x \geq 0.22$ (10 wt.% Mo). Below 10 wt.% Mo concentration, the resulting thin film alloys exhibited a mixed α-γ uranium phase composition.

**Keywords:** Magnetron Sputtering, Uranium-molybdenum alloys, Polycrystalline thin films, Single-crystal thin films, Structure, X-ray Diffraction, Focus Ion Beam, Scanning Electron Microscopy.

## 1. Introduction

Uranium is a fundamental component of most nuclear fuels. Whilst early fission reactors used natural uranium metal as the fuel material, uranium dioxide ($UO_2$) quickly became the predominant nuclear fuel in later reactors. For the next generation of nuclear reactors, the use of metallic fuels has once again become an option because they offer properties that promise to improve the fuel performance in terms of sustainability, reliability, and safety. Due to expectations that they will perform better under irradiation and thermal cycling, uranium alloys with Nb, Mo or Zr have recently been investigated as alternatives to pure uranium or $UO_2$. Such alloy materials can be successfully produced in thin film form using magnetron sputtering technology. It has previously been shown that binary alloy thin films can form nanocrystalline structures with grain size < 10 nm [1]. Such samples are more easily handled and transported than bulk alloy samples due to the negligible radioactivity exhibited. In addition, this technique allows for the stabilisation of the high-temperature phases of the alloy at low temperatures [1], a characteristic which is crucial for development of potential next generation fuel materials such as high-temperature (high-$T$) γ-phase U-alloys. Nowadays, the modern combinatorial approach to thin films synthesis and highly-efficient characterization tools provide composition-structure-property relations [2], hence to indentify compositions with the most desired properties.

Metallic uranium has three crystalline phases below its melting point. The room temperature (RT) α-phase (α-U) is orthorhombic with space group *Cmcm*, unit cell parameters $a = 2.854$ Å, $b = 5.869$ Å, and $c = 4.955$ Å [3-4]. In fuels this phase shows very poor operational stability [5-7], related to highly anisotropic thermal expansion (thermal expansion coefficients for α-U single crystal over the temperature range $T = 25-650°C$ are: $α_{[100]} = 23.53×10^{-6}$ °C$^{-1}$, $α_{[010]} = 1.16×10^{-6}$ °C$^{-1}$ and $α_{[001]} = 19.38×10^{-6}$ °C$^{-1}$ [8]). It also exhibits poor dimensional stability during irradiation, making it undesirable as a fuel. Irradiation growth refers to the change in uranium shape at constant volume without any external stress application. A single crystal of α-U during irradiation grows in the [010] direction, shrinks in the [100] direction and remains unchanged in the [001] direction [9]. In a polycrystalline sample the extent of dimensional changes depends on the degree of preferred orientation in the specimen. The tetragonal β-phase of uranium (β-U) exists between 668 and 775°C, the space group is *P42/mnm*, with unit cell parameters $a = 5.656A$ Å and $b = c = 10.759$ Å [10], whilst the high-temperature ($T > 775°C$) uranium γ-phase (γ-U) is a body-centered cubic (*bcc*) structure with space group *Im3m* and cell parameter $a = 3.524$ Å. This later phase is more resistant to irradiation effects [11] than the α-phase and exhibits isotropic thermal expansion properties. Unfortunately, in pure U this phase cannot be preserved to RT, which presents a significant technical limitation. Conceptually, a solution to this problem would be to stabilise the γ-phase down to RT by alloying with Nb, Zr or Mo [11-12]. Accordingly, in this paper we explore the binary alloy phase diagram of both poly- and single- crystal U-Mo thin films.

The bulk U-Mo system has several metastable phases, depending on the Mo content and the cooling rate following solution treatment at high temperatures. In equilibrium, above 560°C [12] the U-Mo system has cubic structure (space group *Im3m*) denoted as a γ-phase, which is a solid solution in the range of 0-40 at.% (0-21 wt.%) Mo. Below 560°C the γ-phase decomposes into an α-U, which contains less than 1.0 at.% Mo and an ordered U$_2$Mo γ'-phase (body-centered tetragonal structure, space group *I4/mmm*). Water-quenched U-alloys containing up to 6.0 at.% (2.5 wt.%) Mo are found to form to an orthorhombic α'-phase [13]. For Mo concentration between 6.0 and 11.2 at.% (2.5-4.8 wt.%) the monoclinic α''-phase was confirmed [14]. For higher alloy concentrations, *i.e.* between 9.0 and 12.0 at.% (5.2 wt.%) Mo, the alloy has a reported double-phase (α'' + γ-like) structure with the α''-phase predominantly. In general, water-quenched (from $T$ within the γ-field) U-Mo alloys containing up to 11.0 at.% (4.7 wt.%) Mo may be called "α-phase" alloys as the structure of metastable phases obtained (α' and α'') differs slightly from that of α-U [15]. Conversely, alloys with more than 11.0 at.% Mo are often called "γ-phase" alloys as the structures are related to the structure of γ-U. Indeed, body-centered tetragonal structures, designated as the γ$^0$-phase (space group *I4/mmm*), are obtained in alloys containing 11.4 to 12.7 at.% (5.0- 5.6 wt.%) Mo [14-17]. The minimum content of Mo required to ensuring a single γ-phase alloy was found to be 13 at.% Mo [18]. In summary, the phase transformations of U-Mo system depending on Mo content follow the sequence: α→α'→α''→γ$^0$→γ.

Bulk U-Mo alloys containing 11.6, 21.7 and 30.5 at.% Mo exhibit superconductivity below $T_{sc} = 2$ K [19]. The same phase transition was found in splat-cooled materials with $T_{sc}$ in the range 1.24 K (for pure U-splat) to 2.11 K (for U-15 at.% Mo) [20]. In terms of magnetic properties, bulk U-Mo alloy samples with compositions 15-30 at.% Mo exhibit a weak Pauli paramagnetism [21]. More detailed investigation of physical properties including superconductivity, magnetism and mechanical properties of U-Mo single crystal samples will be the subject of future report.

In the current study we have investigated (meta) stabilization of the *bcc* γ-phase in poly- and single-crystal $U_{1-x}Mo_x$ alloy thin films formed by *dc* magnetron sputtering for Mo concentrations up to 29 at.% (14 wt.%). By studying thin film samples we hope to improve our fundamental understanding of alloying by producing high purity samples of well controlled chemistry and thereby excluding many of the problems associated with bulk counterparts. Additionally, these samples provide ideal surfaces for corrosion and oxidation studies [22]. Phase analysis was performed using X-ray Diffraction (XRD). Microstructure was determined using Scanning Electron Microscopy (SEM) and Electron Backscatter Diffraction (EBSD), whilst Energy-Dispersive X-ray (EDX) analysis and Auger Electron Spectroscopy (AES) were used to determine the elemental composition. The thickness of the films was directly measured from cross-section analysis using combined Focused Ion Beam (FIB) milling and electron microscopy.

## 2. Experimental details

### 2.1. Fabrication of thin films

Poly- and single-crystal U and U-alloy thin films were grown using a four gun *dc* magnetron sputtering system with an UHV base pressure of $10^{-10}$ mbar, *in situ* RHEED analysis and a substrate heating stage capable of temperatures up to 850°C.

Alloy sputtering was carried out in an argon environment with a pressure of $8.0 \times 10^{-3}$ mbar. Substrates of size 12.0×10.0×1.0/0.5 mm were microscopy glass, single-crystal silicon and epitaxially polished single-crystal sapphire ($Al_2O_3$) of [11.0] orientation plates. Prior to loading into the UHV system, the substrates were cleaned by boiling successively in acetone, propanol and methanol. A 30 nm thick niobium buffer layer was used to seed crystalline growth of single-crystal U-Mo alloys. Nb has a body centred cubic (*bcc*) crystal structure (space group *Im*3*m*) with the lattice parameter $a = 3.303$ Å and it grows epitaxially at elevated temperatures with [110] direction along growth axis. After alloy deposition a protective a capping layer of Nb or Mo was deposited to prevent atmospheric attack. A study of epitaxial [110] Nb films deposited on sapphire show that a stable (2.0 nm thick) layer of $Nb_2O_5$ is formed, which provides effective long-term passivation [23]. Fig. 1 shows schematic representation of polycrystalline and single-crystal U-Mo thin films. Sample synthesis details are summarised in Table 1.

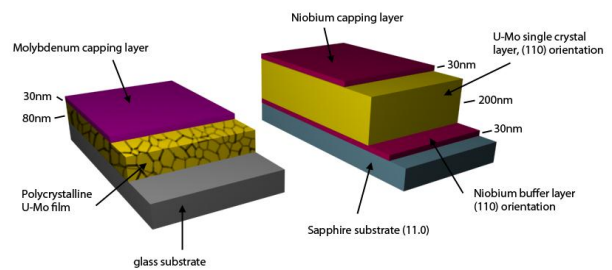

**Fig. 1.** Comparison of polycrystalline (left) and single-crystal (right) U-Mo thin films.

**Table 1.** Summary of sample synthesis.

| Thin film | Substrate | Buffer layer | Capping layer | Crystallinity | $T$ (°C) | Thickness (nm) | Sputtering rate (nm/s) |
|---|---|---|---|---|---|---|---|
| Mo | Si | - | - | P* | RT 400 | 30 | 0.12 |
| U | Si | Mo | Mo | P | RT 400 | 150 | 0.15 |
| U-Mo alloys | glass sapphire | - Nb | Mo Nb | P SC# | 700 | 80 200 | 0.2 |

*Polycrystalline
#Single crystals

*2.2. Phase distribution and microstructure analysis*

For phase analysis, XRD experiments were performed using a Philips X'PERT X-ray diffractometer with Cu-Kα radiation ($\lambda$ = 1.54 Å). The conventional $\theta$-$2\theta$ and $\omega$ scans were measured.

SEM and FIB milling were used to examine the microstructure of the samples produced. The SEM was a Zeiss EVO MA10 fitted with $LaB_6$ electron source and a Digiview 3 high speed camera with associated EBSD instrumentation from AMETEK and the FIB was a FEI FIB Strata 201 focused ion beam system (operating under a background pressure of less than $10^{-5}$ mbar) with gallium source. Prior to EBSD analysis of the U-Mo thin films, the sample surfaces were sputtered under vacuum with an argon-ion gun operating at 5 kV in order to remove the capping Nb layer for EBSD analysis. AES spectra were collected in order to confirm that the surfaces were Nb-free. The diffraction data acquired from EBSD analysis were recorded and processed using OIM™ software, which produced crystallographic orientation and phase maps from predefined surface areas using an automated mapping routine.

*2.3. Composition and thickness measurements*

The thickness of the U-Mo thin films was measured in a DualBeam (DB) instrument (a combined FIB-SEM instrument). The FIB was used to deposit a protective platinum coating followed by sectioning to a depth of 2-3 μm, the thickness of the thin film layer was measured using high magnification SEM (see Fig. 1).

Also in the DB successive layers of material were sputtered from the U-Mo alloy thin film samples with EDX analysis acquired at each step after subtracting the elemental signal from the substrate. This produced an elemental depth profile of the polycrystalline thin film samples. In the case of single-crystal U-Mo thin films, the EDX spectra were collected from the cross-sectional face of FIB cuts through.

Complementary elemental analysis was also performed by AES using an Auger spectrometer (JEOL, model JAMP 30) equipped with a $LaB_6$ electron source, a double-pass cylindrical mirror analyser (CMA) and a differentially pumped ion gun. The surface capping layer and alloy films were sputtered using a 4-5 keV argon-ion beam for 1-3 min, rastered over 2×2 mm area and the AES spectra were subsequently collected to generate depth profiles.

## 3. Results and Discussion

### 3.1. Characterization of polycrystalline U thin films

Polycrystalline α-U thin films were grown on Si substrates at both, RT and $T = 400^{\circ}$C (higher-$T$ improves crystallinity). Previously grown α-U thin films directly on glass substrates (not described in this report) exhibited a large amount of uranium oxide or carbide impurities. In order to synthesise pure α-U thin film, the oxide-free substrates (Si) and buffer layers of Mo were used. The α-U layer was approximately $150.0 \pm 5.0$ nm thick, while the bounding Mo layers were approximately $30.0 \pm 5.0$ nm thick (Fig. 2).

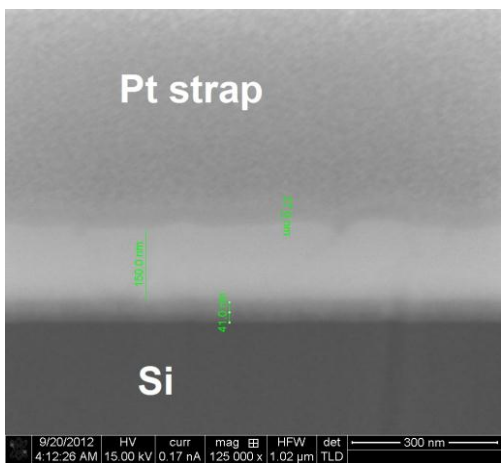

**Fig. 2.** Secondary electron (SE) image of cross-sectional cut through an α-U film.

The XRD pattern acquired from the high-$T$ sample is presented in Fig. 3. It shows polycrystalline α-U and XRD lines of Mo and Si-substrate. The diffracted intensity is plotted against the momentum transfer, $Q = (4\pi/\lambda) \sin\theta$ (Å$^{-1}$), and normalized to unity at the peak of the scattering from the plane (110) of α-U. The α-U (110), (002) and (111) peaks are positioned at 2.439(1) Å$^{-1}$, 2.524(1) Å$^{-1}$ and 2.743(1) Å$^{-1}$, respectively. These reflections can be compared with the wavevectors of the reflections for bulk α-U i.e. (110) at 2.448(1) Å$^{-1}$, (002) at 2.536(1) Å$^{-1}$ and (111) at 2.757(1) Å$^{-1}$ [24]. The peak corresponding to the molybdenum reflection (110) is seen at wavevector transfer of 2.828(1) Å$^{-1}$. No traces of oxide or carbide impurity phase have been detected by XRD.

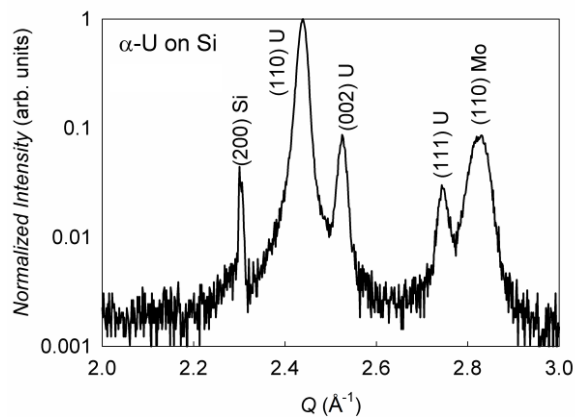

**Fig. 3.** XRD pattern of an α-U film grown on a Si substrate at $T = 400^{\circ}$C.

Analysis and assignment of the peaks in the AES spectra identified elements Mo, O and C to be the predominant elements at the uppermost surface of the film. The recorded oxygen signal was very weak throughout the thickness of the uranium film see Fig. 4, indicating high quality and purity in the grown samples, with only trace levels of oxide.

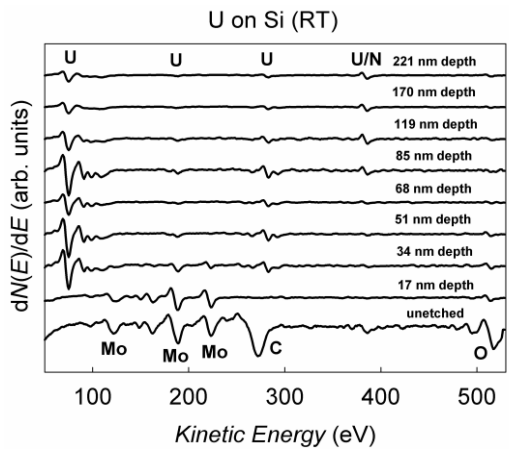

**Fig. 4.** A series of derivative AES spectra displaying a depth profile through U deposited on Si substrate at RT and capped with protective layer of Mo (30 nm thick).

## 3.2. Characterization of polycrystalline $U_{1-x}Mo_x$ thin film

Polycrystalline $U_{1-x}Mo_x$ alloy thin films were grown directly on glass substrates at $T = 700^{\circ}C$ and capped with a protective Mo layer (30 nm thick) at RT. Comparative XRD patterns recorded from different $U_{1-x}Mo_x$ alloy thin films with nominal Mo composition of 10-29 at.% Mo (i.e. $x = 0.10, 0.16, 0.20, 0.26$ and $0.29$) are shown in Fig. 5. The pure uranium γ-phase is suggest the grain size to be less than 5 nm in diameter. Calculated lattice parameters of γ-$U_{1-x}Mo_x$ *bcc* structure are $a = 3.417(1)$ Å for $x = 0.26$ and $a = 3.391(1)$ Å for $x = 0.29$, respectively. For $x < 0.26$ the samples contained both, α- and γ-phases with no evidence for α'-, α''- or $γ^0$-phases reported for bulk U-Mo alloys. Peaks seen at wavevector transfers of $2.824(1)$ Å$^{-1}$ and $3.975(1)$ Å$^{-1}$ were attributed to Mo in the cap and buffer layers, corresponding to only observed for Mo contents $\geq 26$ at.% ($x = 0.29$-$0.26$). The wavevectors of γ-U reflections (110), (200) and (221) for $U_{0.74}Mo_{0.26}$ and $U_{0.71}Mo_{0.29}$ are presented in Table 2. Measurements of rocking curves through the (110) reflection of γ-$U_{1-x}Mo_x$ alloy thin films the reflections (110) and (200), respectively. The traces of $UO_2$ and UC present in the samples are considered to originate from the oxidized Mo capping layer. In order to confirm this statement, the AES spectra for RT and high-$T$ Mo thin films deposited on Si were collected. They showed the presence of oxide through the film. The high-$T$ sample revealed better crystallinity (confirmed by XRD) and weaker oxide signal than the RT one. For the $U_{1-x}Mo_x$ alloy thin film where $x = 0.10$, the XRD lines of $UO_2$ and UC are only present.

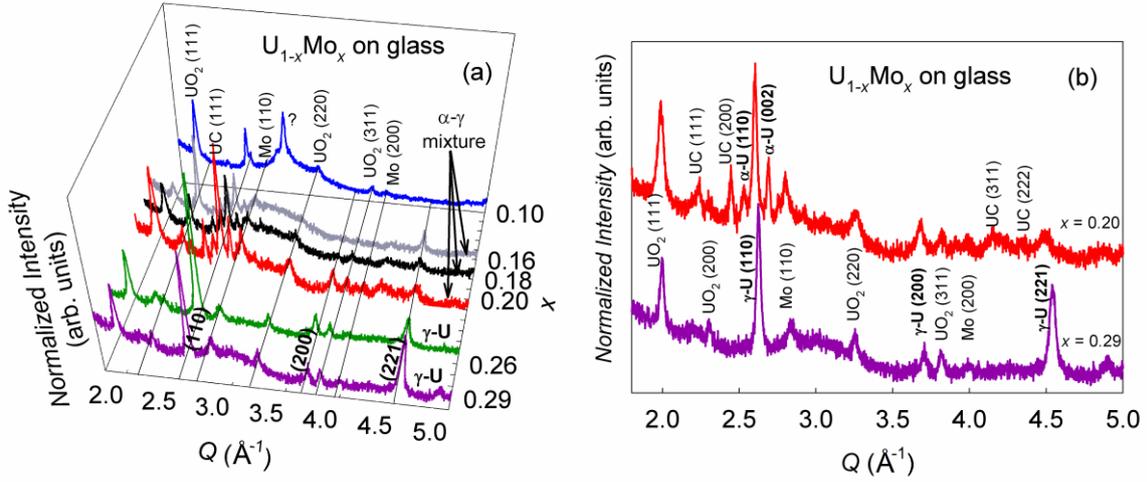

**Fig. 5.** (a) XRD patterns of a series $U_{1-x}Mo_x$ alloy thin films deposited on glass at $T = 700^{\circ}C$, (b) comparison of XRD patterns recorded for pure γ-phase and α- and γ-phase mixed alloys. The Mo metal was used as a capping layer.

**Table. 2.** Wavevector transfer values of the main γ-U reflections for polycrystalline $U_{1-x}Mo_x$ alloy thin films.

| $x$ | $h$ | $k$ | $l$ | $Q$ (Å$^{-1}$) | $2\theta$ (deg.) |
|---|---|---|---|---|---|
|  | 1 | 1 | 0 | 2.623(1) | 37.48 |
| 0.29 | 2 | 0 | 0 | 3.703(1) | 54.12 |
|  | 2 | 2 | 1 | 4.538(1) | 67.58 |
|  | 1 | 1 | 0 | 2.604(1) | 37.33 |
| 0.26 | 2 | 0 | 0 | 3.684(1) | 53.72 |
|  | 2 | 2 | 1 | 4.497(1) | 67.23 |

As it is difficult to distinguish both metals i.e. Mo cap and U-Mo alloy using SEM, the thickness of the layers was separately measured from cross-sectional cut through using a FIB instrument. For each of the samples, the Mo cap was found to be 30.0 ± 5.0 nm thick Fig. 6(a), while the U-Mo alloy layer was 80 ± 5.0 nm thick, see Fig. 6(b). Fig. 7 shows the equivalent EDX plot of U and Mo concentrations for selected $U_{1-x}Mo_x$ alloy thin film. The EDX spectra were collected before and after the removal of the Mo layer using FIB of $Ga^+$ atoms ($I$ = 9 nA). The final composition of the U-Mo alloys was taken as the average of these measurements yielding the values of 71 at.% U and 29 at.% Mo.

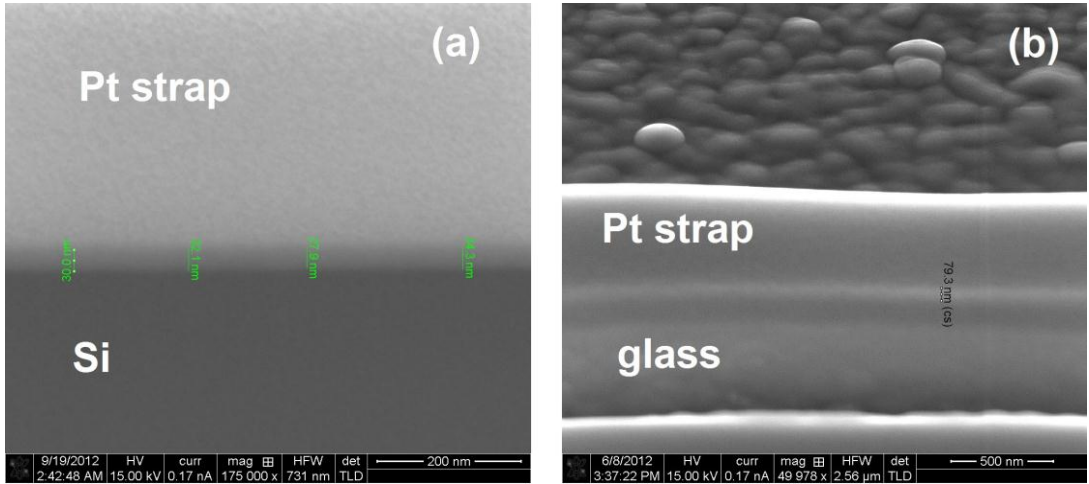

**Fig. 6.** SE images of sections through (a) Mo and (b) uncapped U-Mo alloy thin film sample.

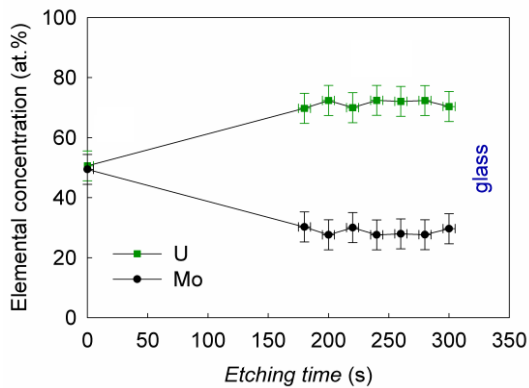

**Fig. 7.** Plot of U and Mo concentrations for polycrystalline $U_{1-x}Mo_x$ alloy thin film deposited at $T = 700^oC$ on glass.

### 3.3. Characterization of single-crystal $U_{1-x}Mo_x$ thin films

Single-crystal $U_{1-x}Mo_x$ alloy thin films were grown and seeded with Nb layers (30 nm thick) on sapphire substrates at $T$ = 700 and $500^oC$. All samples were capped with protective Nb layer (30 nm thick) at RT. The XRD patterns of single-crystal $U_{1-x}Mo_x$ alloy thin films grown at $T = 700^oC$ are shown in Fig. 8. The intensity is normalized to unity at the peak of the scattering recorded from the epitaxial sapphire substrate at $Q$ = 2.642 Å$^{-1}$. A pure uranium γ-phase is observed for 22 at.% Mo ($x = 0.22$), and exhibits a [110] orientation on sapphire when deposited at $T = 700^{\circ}$C. For lower Mo concentrations both, the uranium α- and γ-phases are identified. Small domains of the α-U phase were detected in the γ-U alloy with 20 at.% Mo, exhibiting the [110] orientation, while in the U-alloy with 10 at.% Mo the α-U phase is polycrystalline. The γ-U (110) is observed at 2.622(1) Å$^{-1}$ and 2.616(1) Å$^{-1}$ for $U_{0.78}Mo_{0.22}$ and $U_{0.80}Mo_{0.20}$, respectively. The α-U (110) is located at 2.447(1) Å$^{-1}$ for $U_{0.80}Mo_{0.20}$, which is in agreement with the position of the (110) peak from the high-$T$ polycrystalline α-U deposited on Si. The calculated lattice parameters of γ-$U_{1-x}Mo_x$ bcc structure are therefore $a = 3.399(1)$ Å for $x = 0.22$, $a = 3.407(1)$ Å for $x = 0.20$ and $a = 3.423(1)$ Å for $x = 0.10$, respectively. There is a linear dependence on Mo concentration, that can be described by the equation $a = 3.4421 - 0.1871x$, where $x$ is at.% Mo. However, these values are much lower in comparison to lattice parameters of e.g. hot rolled cast $U_{1-x}Mo_x$ alloys [12], where $a = 3.414$ Å for $x = 0.22$ and $a = 3.431$ Å for $x = 0.20$. Thin films show different lattice behaviour than bulk materials because the thin film is clamped to the substrate and the lattice structure is modulated by the stress from the substrate. The broad hump on the high angle side of the substrate peak reflection ($Q = 2.691$ Å$^{-1}$) corresponds to (110) reflection from Nb. No traces of $UO_2$, UC or other impurity phase have been detected by XRD, suggesting efficient fabrication and subsequent integrity of the capping Nb layer. In order to check the crystallographic quality of the samples, the ω scan was performed by rocking the sample through the Bragg position, while keeping the detector in a fixed position so that the scattering vector runs parallel to $Q_x$, where $Q_x$ is the $x$ component of the scattering vector $Q$. This component can be obtained from the ω and θ angles using $Q_x = (4\pi/\lambda) \sin\theta \sin(\omega - \theta)$ (Å$^{-1}$) [25-26]. Thus, the rocking curves through the (110) reflection of γ-$U_{1-x}Mo_x$ single-crystal alloy thin films were measured, see Fig. 9.

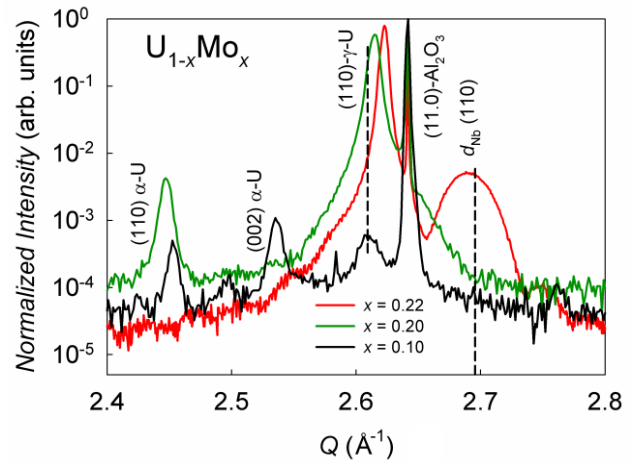

**Fig. 8.** Comparison of the XRD patterns close to the sapphire (11.0) peak for single-crystal $U_{1-x}Mo_x$ alloy thin films deposited at $T = 700^oC$. Nb metal was used as a capping and buffer layer.

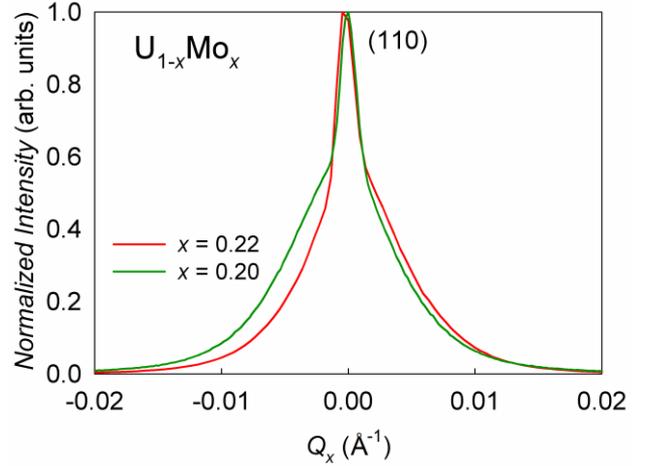

**Fig. 9.** ω scans through the (110) reflection of single-crystal γ-$U_{1-x}Mo_x$ thin films deposited at $T = 700^{\circ}$C.

Two components are observed in the rocking curves, a narrow peak with FWHM ≈ 0.03 deg. and a broader peak with FWHM ≈ 0.2 deg., indicating that the U-Mo films are comprised of abundant low-angle tilt boundaries. The rocking curve width indicates coherency across the film-substrate interface and also the capping layer. A significant portion of the U-Mo film maintains a high degree of coherency, leading to a narrow rocking curve component (FWHM ≈ 10$^{-3}$ deg. for a true epitaxial thin film), while the balance of the film becomes incommensurate, resulting in a broad rocking curve component [27]. For $x = 0.10$ only one broad rocking curve component (FWHM ≈ 3.0 deg.) was observed.

XRD patterns collected for $U_{0.78}Mo_{0.22}$ single-crystal thin films grown at $T < 700^{\circ}$C (Fig. 10) show the mixed α- and γ-U phase, with predominance of the cubic phase. The calculated lattice parameter of γ-U bcc structure obtained at $T = 500^{\circ}$C is $a = 3.406(1)$ Å, which is higher (by about 0.2 %) than that for $U_{0.78}Mo_{0.22}$ synthesised at $T = 700^{o}C$. The boundary of the binary phase diagram for bulk U-Mo system (in this range of Mo composition) between pure γ-U and a mixed phase is at $560^{o}C$ [12]. Fig. 10 shows clearly a similar behaviour in the U-Mo thin films and expanded U-Mo lattice parameter in the γ-phase.

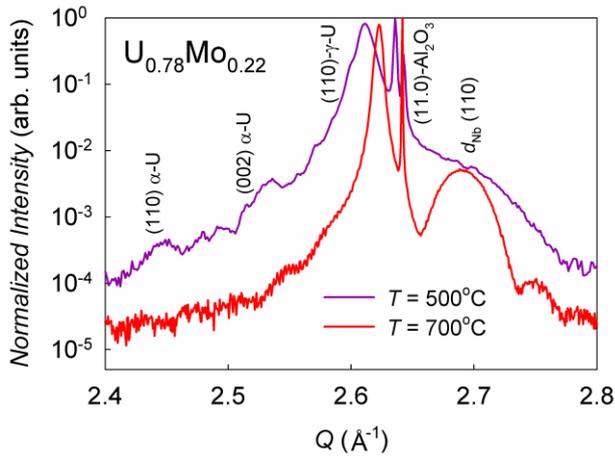

**Fig. 10.** Comparison of the XRD patterns close to the sapphire (11.0) peak for $U_{0.78}Mo_{0.22}$ single-crystals deposited at different temperatures. Nb metal was used as a capping and buffer layer.

EBSD is an extremely surface sensitive diffraction mapping technique with a sampling depth of between 5.0 and 10.0 nm depending on the material being analysed. Consequently it is very well suited to investigation the surface of thin films which may have initially grown epitaxially on a selected substrate but later exhibited more randomised crystal growth. In the current case the data indicates that for $U_{0.78}Mo_{0.22}$ sample there was an almost perfect γ-U single crystal with the [110] orientation (Fig. 11). The confidence index for γ-U is very high ($\approx 1$), while the recorded domains of misorientation (< 2 deg.), displaying as green dots in Fig. 11(b) were considered to be responsible for the broad rocking curve component observed in the XRD data. No α-phase was detected, which was anticipated from the XRD results.

The U-Mo layer thickness was measured to be around $200.0 \pm 5.0$ nm, while the thickness of Nb capping/buffer layers is about 30-50 nm. In order to verify the composition of the alloys, EDX spectra were acquired from the cross-sectional cut of the sample (Fig. 12). The final composition of the U-Mo alloys was taken as the average of numerous (> 10) measurements for each sample. These yield more precise values of the alloy composition than that for polycrystalline samples. For comparison, EDX spectra have also been collected from the surface of the sample (Fig. 13).

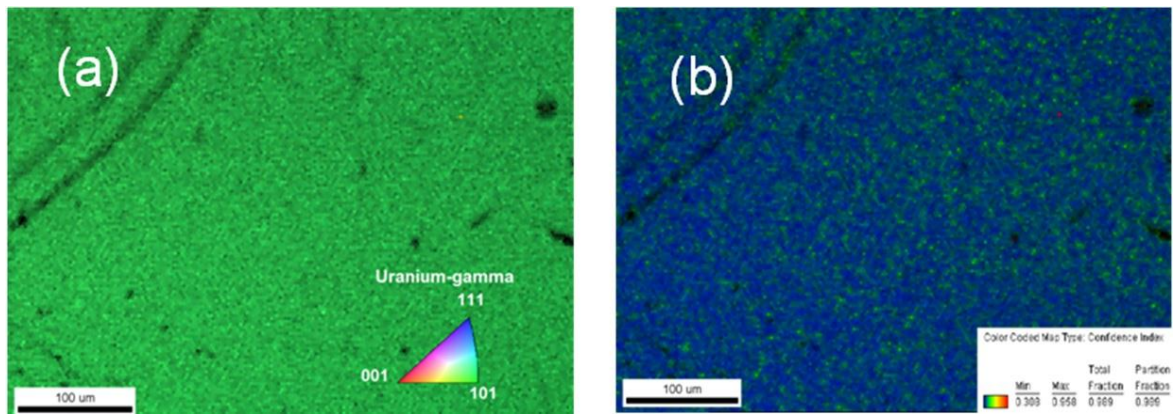

**Fig. 11.** The Inverse Pole Figure EBSD maps of $U_{0.78}Mo_{0.22}$ showing (a) the orientation of the single crystal and (b) confidence index for the γ-U phase.

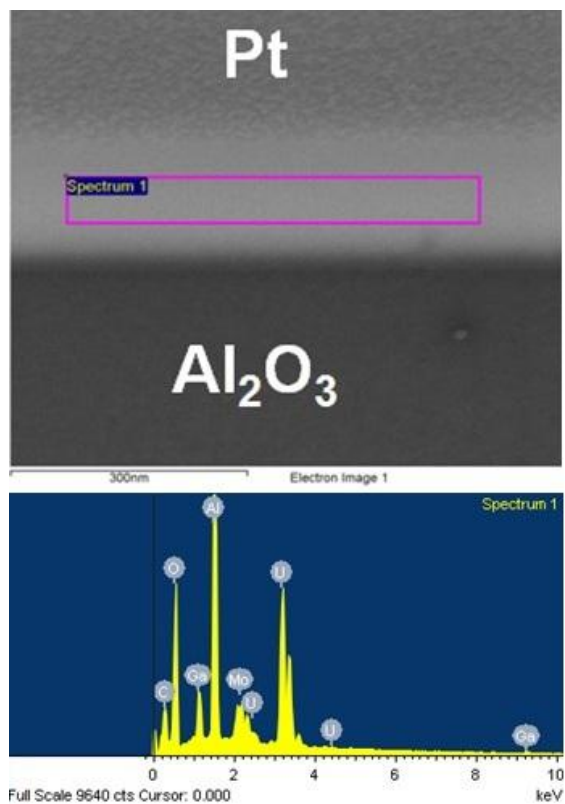

**Fig. 12.** SE image of the cross-sectional cut through U-Mo alloy thin film (top), the square indicates the region where the EDX spectrum (bottom) has been acquired.

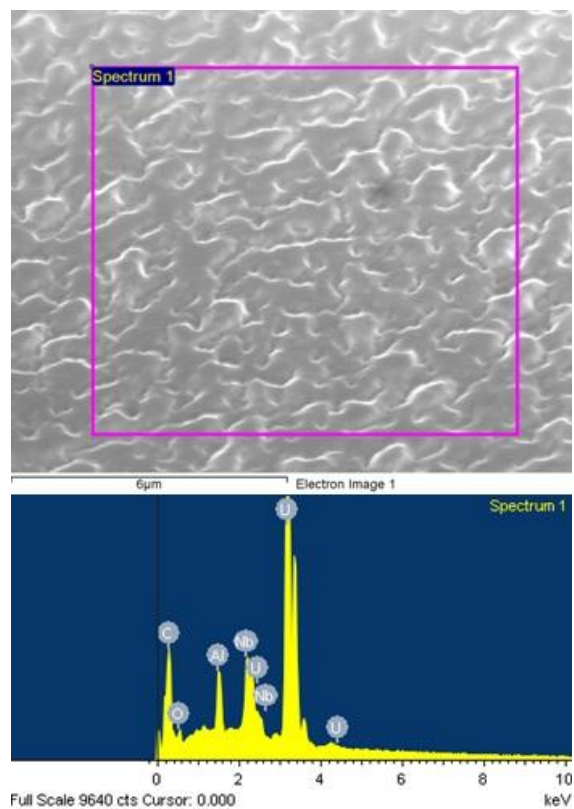

**Fig. 13.** SE image of the surface of U-Mo alloy thin film (top), the square indicates the region where the EDX spectrum (bottom) has been acquired.

## 4. Conclusions

The key in developing an effective alloy fuel is in stabilizing a wholly γ-phase material. To date the most effective uranium alloy identified is U- 20 at.% (9.1 wt.%) Mo. It possesses high radiation resistance, low swelling under irradiation, high strength and ductility and good resistance to corrosion [18]. However, a potential drawback for this alloy composition is that the fissile isotope is sufficiently diluted that enrichment would be needed for this alloy to be viable as a fuel, thereby significantly increasing its cost. Following this, other studies have explored complex metallurgical preparations such as heating/cooling in an induction furnace [13], splat cooling [20] and *etc*. were applied to stabilise the γ-phase U-Mo alloys with the lowest possible Mo content. We have shown that high quality poly- and single-crystal U-Mo alloy thin films with various thicknesses can be grown using an UHV sputtering facility at $T = 700^{o}C$. These results obtained for polycrystalline and single-crystal $U_{1-x}Mo_{x}$ are summarized in Table 3. The γ-U phase stability difference observed may relate to partitioning of Mo at grain boundaries in the polycrystalline films, such that a portion of the sputtered Mo is not efficiently participating in alloying. Future work will utilise high resolution transmission electron microscopy (TEM) to resolve the structure and elemental distribution within different polycrystalline U-Mo samples.

We have determined that the ideal base for growing single-crystal of U-Mo alloy thin films was a niobium (110) buffer layer grown on sapphire (11.0). The lattice parameter calculated for Nb single crystal is $a = 3.301(1)$ Å, while $a = 3.399(1)$ Å for $U_{0.78}Mo_{0.22}$, resulting in a lattice mismatch of approximately 3 %. In addition, the RHEED analysis of the Nb capping layer revealed epitaxial growth of Nb metal on the U-Mo thin film. The well-defined narrow component of the rocking curve indicates that a significant portion of the sample has a high degree of coherency across the film-substrate interface. The results clearly show that it is possible to grow high quality single crystals of γ-U-Mo alloy by physical vapour deposition (PVD), such an approach provides a highly useful source of material for on-going investigations of thermal cycling, irradiation stability and oxide formation.

**Table 3.** Summary of results for polycrystalline and epitaxial $U_{1-x}Mo_x$ thin films.

| Polycrystalline samples | Single crystal samples |
| --- | --- |
| γ-phase for $x \geq 26$ at.% Mo | γ-phase for $x \geq 22$ at.% Mo |
| One component rocking curve with FWHM ≈ 30 deg. | Two component rocking curves with FWHM ≈ 0.03 and 0.2 deg. |
| α-γ-mixture for $x \leq 20$ at.% Mo | No impurity phase |
| $UO_2$, UC present | |
| Grains < 5 nm (no EBSD mapping) | $U_{0.78}Mo_{0.22}$ – perfect γ-phase single crystal (misorientations < 2 deg.) |

**Acknowledgements**

European Commission, People Marie Curie Actions - Intra-European Fellowships (IEF) for Career Development, Call: FP7-PEOPLE-2011-IEF, Project No: 300171, Project Acronym: URALLOY.